\documentclass[twoside,journey]{IEEEtran}

\usepackage{makecell}
\usepackage{hyperref}
\usepackage{array}
\usepackage{caption2}
\usepackage{graphicx,amssymb,amsmath}
\usepackage{multicol}
\usepackage[noadjust]{cite}
\usepackage{setspace}
\usepackage{subfigure}
\usepackage{graphicx}
\usepackage{float}
\usepackage{url}
\usepackage{stfloats}
\usepackage{amsthm,pifont}
\usepackage{flushend}
\usepackage{cases,subeqnarray}
\usepackage{bm,multirow,bigstrut}
\usepackage{amsmath, amsthm, amssymb}
\usepackage{textcomp}
\usepackage{latexsym,bm}
\usepackage{booktabs}
\usepackage{xcolor}
\usepackage{mathtools}
\usepackage{dsfont}
\usepackage{extarrows}
\usepackage{epsfig}
\usepackage{epsfig}
\usepackage{epstopdf}
\usepackage[noend]{algpseudocode}
\usepackage{algorithmicx,algorithm}

\theoremstyle{plain}

\theoremstyle{plain}

\usepackage{amsmath}

\IEEEoverridecommandlockouts
\begin{document}
\title{Fusion of Mixture of Experts and Generative Artificial Intelligence in Mobile Edge Metaverse}

\author{Guangyuan Liu, Hongyang Du, Dusit Niyato,~\IEEEmembership{Fellow,~IEEE}, Jiawen~Kang, Zehui~Xiong, Abbas Jamalipour,~\IEEEmembership{Fellow,~IEEE}, Shiwen Mao,~\IEEEmembership{Fellow,~IEEE}, and Dong In Kim,~\IEEEmembership{Fellow,~IEEE}
\thanks{G.~Liu, and H.~Du are with the College of Computing and Data Science, the Energy Research Institute @ NTU, Interdisciplinary Graduate Program, Nanyang
Technological University, Singapore (e-mail: liug0022@e.ntu.edu.sg, hongyang001@e.ntu.edu.sg).}
\thanks{D. Niyato is with the College of Computing and Data Science, Nanyang Technological University, Singapore (e-mail: dniyato@ntu.edu.sg).}
\thanks{J. Kang is with the School of Automation, Guangdong University of Technology, China. (e-mail: kavinkang@gdut.edu.cn).}
\thanks{Z. Xiong is with the Pillar of Information Systems Technology and Design, Singapore University of Technology and Design, Singapore (e-mail:
zehui\_xiong@sutd.edu.sg).}
\thanks{S. Mao is with the Department of Electrical and Computer Engineering, Auburn University, Auburn, USA (e-mail: smao@ieee.org)}
\thanks{A. Jamalipour is with the School of Electrical and Computer Engineering, University of Sydney, Australia (e-mail: a.jamalipour@ieee.org).}
\thanks{D. I. Kim is with the Department of Electrical and Computer Engineering, Sungkyunkwan University, South Korea (e-mail: dongin@skku.edu).}
}
\maketitle
%----------------------------abstract----------------------------
\vspace{-1cm}
\begin{abstract}
In the digital transformation era, Metaverse offers a fusion of virtual reality (VR), augmented reality (AR), and web technologies to create immersive digital experiences. However, the evolution of the Metaverse is slowed down by the challenges of content creation, scalability, and dynamic user interaction. Our study investigates an  integration of Mixture of Experts (MoE) models with Generative Artificial Intelligence (GAI) for mobile edge computing to revolutionize content creation and interaction in the Metaverse. Specifically, we harness an MoE model’s ability to efficiently manage complex data and complex tasks by dynamically selecting the most relevant experts running various sub-models to enhance the capabilities of GAI. We then present a novel framework that improves video content generation quality and consistency, and demonstrate its application through case studies. Our findings underscore the efficacy of MoE and GAI integration to redefine virtual experiences by offering a scalable, efficient pathway to harvest the Metaverse's full potential.
\end{abstract}


\section{Introduction}

With the advancements in Virtual Reality (VR) and Augmented Reality (AR), the Metaverse is redefining the landscape of digital interaction and user engagement as an emerging paradigm that blends the physical and digital realms. This virtual universe is characterized by its comprehensive integration of Web technologies and extended reality (XR), including VR, mixed reality, and AR, offering an unparalleled immersive experience. However, the creation and development of the Metaverse still faces significant challenges. Traditional creation methods rely heavily on manual creation which are time-intensive and costly. Such static services often fail to create environments that can adaptively respond to changes and user interactions, thereby reduce the immersive experience in the Metaverse~\cite{xu2023unleashing}.

To overcome these challenges, Generative AI (GAI) has been introduced for Metaverse content creation, aiming to enhance the immersive experiences within these virtual spaces. GAI, characterized by its ability to generate content either entirely or partially through AI, stands as a transformative tool in the Metaverse, enabling unprecedented levels of realism and automating complex processes~\cite{qin2023empowering}. This technology not only augments our existing digital world but also paves the way for innovative virtual applications and services.

The potential of GAI in revolutionizing the Metaverse is huge. It can dramatically accelerate and diversify the creation of virtual content, ranging from realistic landscapes to interactive characters and narratives. This capability is crucial in transforming and adapting existing services to new, more immersive contexts. For instance, GAI can convert 2D images into 3D models or adapt real-world audio into immersive soundscapes, thus enriching the Metaverse experience~\cite{qin2023empowering}. Additionally, GAI's roles extend to personalization and adaptive narratives. By referring to user behavior and preferences, GAI can modify storylines and environments in real-time, ensuring that each user's journey within the Metaverse is unique and engaging~\cite{murala2023artificial}.

 However, a significant challenge arises when GAI facing complex prompts or input involving multiple subjects. In such scenarios, there is a tendency for GAI to lose focus on subsequent subjects, diminishing the depth and continuity of the generated content. This issue is particularly pronounced in mobile edge environments, where computational resources are distributed across edge devices. The resultant content often lacks coherence, with later subjects receiving inadequate attention, leading to a fragmented user experience that undermines the immersive quality of the Metaverse. This not only detracts from the realism and engagement intended by these virtual environments but also limits the potential for personalized and adaptive narratives that respond effectively to user interactions.

On the other hand, integrating Mixture of Experts (MoE) with GAI represents a significant advancement in the development of the Metaverse, particularly within mobile edge computing. MoE's ability to handle large-scale and intricate data makes them ideal for addressing the challenge of inconsistency. Selectively activating neural network parts for specific tasks enables more dynamic and responsive virtual environments in the Metaverse.

Utilizing MoE, we propose a novel framework designed to enhance the attention and responsiveness of GAI within mobile edge computing scenarios. Our framework specifically targets the optimization of GAI's content generation process, ensuring balanced attention across all subjects within complex prompts and consistency in the generated content. The contributions of this paper are summarized as follows:
\begin{itemize}
\item We delve into the roles of MoE and GAI, their specific applications in the Metaverse, and the implications of their integration for the future of virtual environments and applications.
\item We explore the integration of MoE and GAI in the Metaverse, investigating their collaborative potential in creating immersive, interactive, and adaptive virtual spaces.
\item We propose a novel framework combining MoE with GAI for enhanced video generation in the Metaverse, optimized for mobile edge computing. The outcomes highlight significant advancements in video quality and consistency, outperforming traditional content generation methods. Our framework's capability is demonstrated to efficiently manage and scale up dynamic video creation, promising a transformative impact on user experiences within the Metaverse.
\end{itemize}
\section{Mobile Edge Metaverse Services}
\begin{figure*}[t!]
\centerline{\includegraphics[width=1\textwidth]{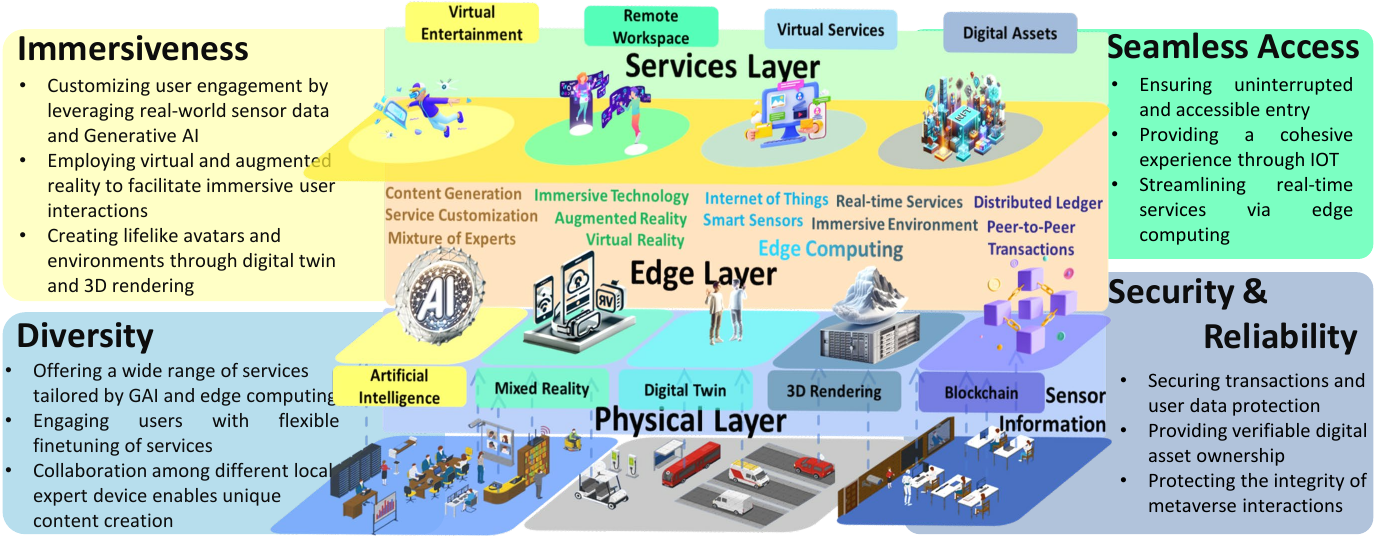}}
    \caption{The mobile edge Metaverse architecture includes three layers: The Physical layers where the sensor can collect information from the real world. The Edge layer then uses the collected data to generate different assets, such as images, video, and audio, to alter existing content to form services. The Services layer forms all assets together for a completed and customized service experience.}
    \label{fig:Metaverse}
\end{figure*}

\subsection{Mobile Edge Metaverse Architecture}
In the emergent landscape of the mobile edge Metaverse, the fusion of state-of-the-art technologies is imperative for the creation of a deeply immersive virtual ecosystem. Recognized as the potential successor to the mobile Internet, the Metaverse concept has surged in prominence and relevance, primarily due to its transformative capacity in reshaping service delivery across diverse sectors including healthcare, education, entertainment, e-commerce, and industrial domains~\cite{dhelim2022edge,xu2022full}. Nonetheless, harvesting this potential and ensuring widespread accessibility necessitate overcoming substantial implementation hurdles, particularly in communication, networking, and computational domains.

Central to the architecture of the Metaverse is its foundation on a harmonious blend of immersive and common virtual ecosystems, navigable through user avatars. This architecture is significantly enhanced by technologies including XR, and edge computing~\cite{xu2022full}, which effectively dissolve the barriers between physical and digital realms. These innovations empower users to engage with the virtual world in a physically interactive manner, transcending the limitations of traditional 2D interfaces and fostering a more immersive experience~\cite{dhelim2022edge}. 

Mobile edge networks are an integral part in addressing the high computational demands, such as rendering intricate 3D virtual worlds and managing AI-intensive avatars, particularly on resource-constrained edge devices. Furthermore, the transition from centralized to distributed or decentralized data architectures across the ``Internet of Everything" will enable actualizing the Metaverse within mobile edge networks~\cite{xu2022full}.

As illustrated in Figure~\ref{fig:Metaverse}, the four main foundational pillars of the mobile edge Metaverse Architecture are Immersiveness, Diversity, Seamless Access, and Security \& Reliability. At its core, the architecture integrates real-world sensory data with GAI and cutting-edge VR and AR technologies to craft lifelike avatars and environments for better Immersiveness. It emphasizes Diversity by offering a broad spectrum of services fine-tuned by GAI and edge computing, alongside fostering unique content creation through collaboration among various local expert devices. Seamless Access is achieved by ensuring uninterrupted entry and a cohesive experience powered by Internet of Things (IoT) technologies, further streamlined by edge computing for real-time services. Lastly, Security \& Reliability are critical for safeguarding transactions, protecting user data, ensuring verifiable digital asset ownership, and maintaining the integrity of Metaverse interactions.

As we navigate toward the holistic realization of the Metaverse, the harmonization of edge computing with other advanced technologies becomes increasingly central. This orchestrated approach deftly addresses the intricate computational and networking challenges of sculpting a truly immersive and interactive virtual world~\cite{dhelim2022edge}. Moreover, within these technological advancements, GAI emerges as a dynamic force, especially within the mobile edge Metaverse. Capitalizing on the strengths of edge computing, GAI operates closer to users, fostering real-time, context-aware content generation and interactions~\cite{qin2023empowering}. Such synergy of GAI with edge computing not only bolsters the responsiveness and adaptability of virtual environments but also ushers in an era of more dynamic and innovative experiences within the mobile edge Metaverse. 
\subsection{Generative AI in the Mobile Edge Metaverse}

Operating at the edge, GAI capitalizes on the computational capabilities and reduced latency provided by edge computing to support mobile Metaverse, thereby ensuring real-time content generation and interaction that are both contextually aware and intensely personalized. The primary advantage of GAI in the mobile edge Metaverse lies in its ability of leveraging advanced models including Transformers, Generative Adversarial Networks (GANs), Normalized Flows, Variational AutoEncoders (VAEs), and diffusion models for content generation. These models harness the power of mobile edge computing to bring in real-time, context-aware, and personalized digital content to users, significantly enriching the Metaverse experience. With a proximity to data sources, edge enabled GAI efficiently crafts intricate landscapes, detailed objects, and coherent environments, overcoming traditional barriers in virtual world building~\cite{qin2023empowering}. Furthermore, operating within the proximity of users, GAI ensures that these creative processes are not only swift but also highly responsive to user interactions, thus enhancing the overall user experience.

In the domain of enhancing character design and interaction, GANs excel in sensing and generating high-quality images and videos, creating realistic virtual environments and avatars. Normalized Flows and VAEs are pivotal in image and audio generation, providing the ability to model complex distributions for creating diverse and rich virtual contents. Diffusion models, especially noted for their prowess in image and video generation, offer a robust framework for creating detailed and coherent visual environments. These models empower users to design avatars that are not only visually unique but also embedded with distinct behaviors and personalities. This strengthens the connection between users and their virtual representations, making interactions within the Metaverse more engaging and vivid. Further, GAI advances the interactivity within these spaces by introducing responsive Non-Player Characters (NPCs) and dynamic elements, such as evolving puzzles, thereby enriching the social and exploratory facets of the Metaverse~\cite{lv2023generative}.

On the other hand, GAI's proficiency in storyline generation and adaptation is crucial in the mobile edge Metaverse. Transformers are well known for its effectiveness in natural language processing, empowering Large Language Models (LLM) to understand and generate human-like text, and making them ideal for applications such as chatbots and virtual assistants within the Metaverse. It meticulously analyzes user interactions and preferences, tailoring experiences and storylines in real-time. This ensures that each user's journey within the Metaverse is unique, adaptive, and immersive~\cite{murala2023artificial}. By processing this data at the edge devices, GAI guarantees that these personalized narratives are delivered swiftly, maintaining the narrative flow and engagement without perceptible delays.

As for facilitating logic content generation in game design and development, GAI's procedural content generation ensures a continuously fresh and engaging experience. By autonomously creating levels, challenges, and puzzles, it maintains a high level of novelty and challenge, catering to a diverse spectrum of preferences and skill levels. This not only keeps the Metaverse vibrant but also sustains user interest over prolonged periods.

However, GAI models face challenges when scaled up to handle the vast and dynamic nature of the Metaverse. The complexity and size of these models often necessitate significant computational resources, which can be a limiting factor when deploying them on user devices within the mobile edge Metaverse. To address these challenges and harness the full potential of GAI in the Metaverse, the integration of MoE emerges as a strategic approach. MoE offers a way to manage the complexity of GAI models by dividing tasks among multiple specialized models (i.e., experts) and selectively activating them based on the task at hand. This not only reduces the computational burden on individual devices but also allows for the dynamic and efficient generation of content, tailored to the specific needs and contexts of users.

\section{Mixture of AI Expert in Mobile Edge Metaverse}

\begin{figure*}[t!]
\centerline{\includegraphics[width=0.95\textwidth]{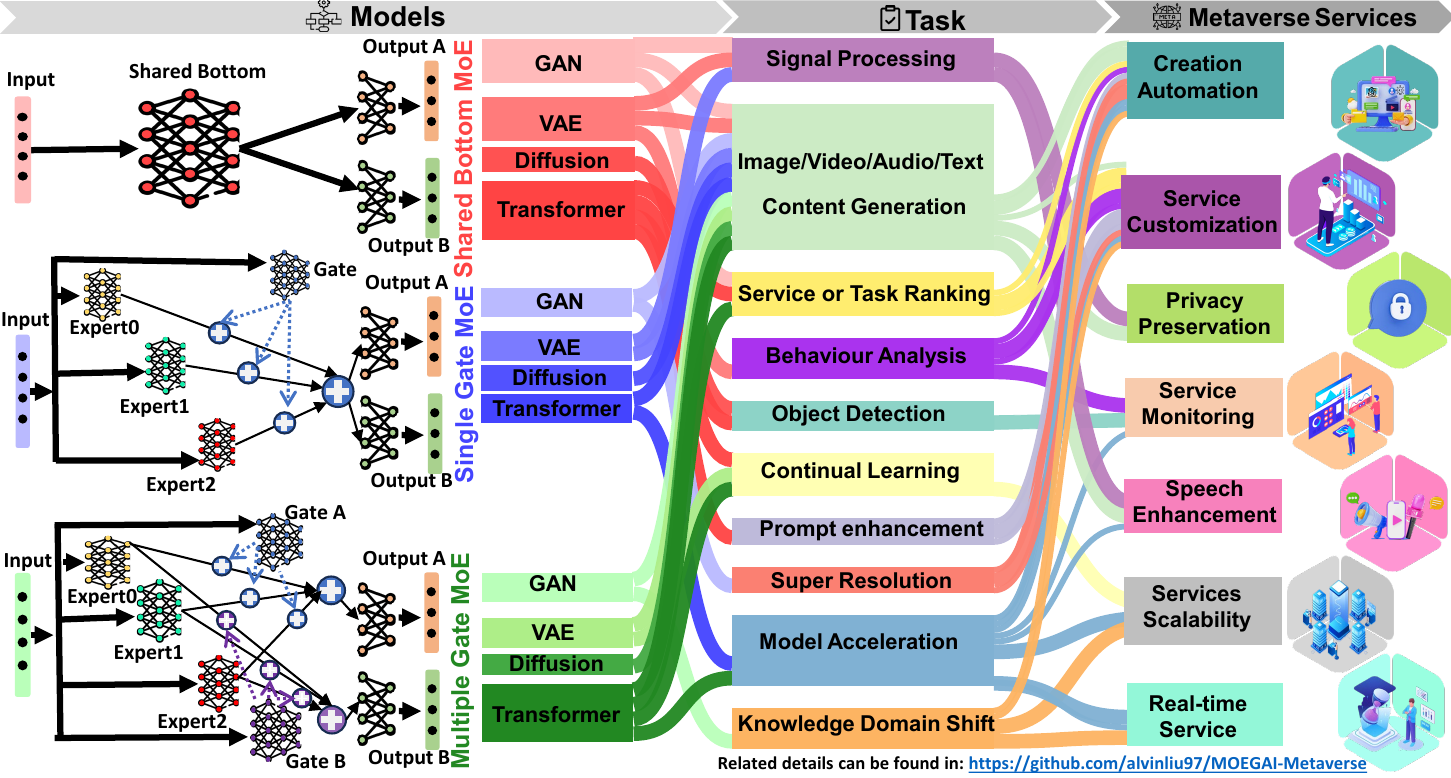}}
    \caption{Applications of MoE in GAI for the Metaverse. For example, the authors in ~\cite{9980170} proposed a multi-gate MoE framework that employs a GAN architecture, where the generator is a speech enhancement network and the discriminator is a speech quality assessment network. This discriminator predicts quality metrics that guide the speech enhancement network to improve the quality of speech by minimizing the discrepancy between enhanced speech and clean speech.}
    \label{fig:category&applications}
\end{figure*}

This section embarks on a detailed exploration of MoE's sophisticated mechanisms and its profound synergy with both Discriminative AI and GAI. By dissecting various MoE structures and their specific applications, we illuminate how this integration not only addresses the aforementioned challenges but also dynamically shapes the evolutionary trajectory of virtual environments, making them more responsive, inclusive, and boundlessly creative.
\subsection{Mixture of Experts: Detailed Exploration}

The MoE model stands as a significant innovation in the field of ensemble learning, particularly within neural network applications. Unlike traditional deep learning models that engage the entire neural network for each input, MoE introduces a more sustainable and efficient approach. It achieves this by selectively activating only certain portions of the neural network based on the input, to produce the relevant output, thereby significantly expanding the model's parameters without a corresponding increase in computational demands.

Central to the MoE's operational principle is the segmentation of the modelling process into distinct sub-tasks. Each sub-task is managed by a dedicated expert model, trained to specialize in certain aspects of the problem. Complementing these expert models, MoE also incorporates a Gating Model. The gating model is calibrated in conjunction with the expert models during training and functions to distinguish which expert model is the most suitable one for a given input. It intelligently integrates their predictions, ensuring a cohesive output that leverages specialized knowledge from multiple domains~\cite{shazeer2017outrageously}.

The versatility of the MoE technique is further exemplified by its adaptability to various model types, including but not limited to Neural Network-based Expert and Gating Models. As illustrated in Figure~\ref{fig:category&applications}, three notable variants of the MoE model are listed below:
\begin{itemize}
    \item \textbf{Shared-bottom model:} The Shared-bottom model integrates a common lower network layer with task-specific layers, facilitating knowledge transfer between related tasks while reducing the amount of parameters. However, its performance might be suboptimal for highly distinct tasks.
    \item \textbf{Single-gate MoE model:} The Single-gate MoE model utilizes a single gating mechanism to determine the appropriate expert network for processing each input, offering a more task-specific approach. 
    \item \textbf{Multi-gate MoE mode:} The Multi-gate MoE model extends this concept by introducing multiple gates, each dedicated to a specific task, thereby allowing for a higher degree of task-specific customization and precise expert selection~\cite{ma2018modeling}.
\end{itemize}

Next, we delve into the realm of Discriminative AI, demonstrating how MoE's adaptive and specialized approach significantly strengthens the capabilities of traditional AI systems.
\subsection{Mixture of Discriminative AI Expert}
The foundational strength of MoE in Discriminative AI is its compatibility with a diverse range of expert architectures. This includes Support Vector Machines (SVMs), Gaussian Processes, Dirichlet Processes, and deep neural networks~\cite{shazeer2017outrageously}. For instance, MoE in SVMs enables effective handling of classification and regression tasks, especially in high-dimensional data spaces. Utilizing Gaussian Processes as experts to enhance the Discriminative AI model's ability to manage uncertainty and make probabilistic predictions is crucial for applications including regression and time series analysis. The Dirichlet Processes can also be employed to allow dynamic adaptation in the number of components based on the data to bolster the clustering and density estimation tasks. Moreover, incorporating deep learning architectures as experts significantly enhances the MoE model's handling of complex, hierarchical data structures, as seen in image and speech recognition tasks~\cite{shazeer2017outrageously}.

The configurational flexibility of MoE models further extends their applicability across complex problems. Hierarchical MoE models organize experts in layers, each addressing different levels or aspects of the problem, enabling structured information processing~\cite{shazeer2017outrageously}. Additionally, the capability to sequentially add experts allows the MoE model to continually evolve and adapt to new data or changing environments, proving invaluable in dynamic systems where task or data nature may shift over time.

Another key advantage of MoE contributing to Discriminative AI is computational efficiency. By activating only a subset of networks (the relevant experts) for a given input, MoE introduces sparsity in neural networks, reducing the computational load~\cite{ma2018modeling,rajbhandari2022deepspeed,fan2022m3vit}. This targeted activation makes MoE models particularly suitable for large-scale and complex AI applications. Furthermore, MoE's ability to decompose complex problems into simpler sub-problems, each handled by a specific expert, not only enhances the model's accuracy but also streamlines the computational process~\cite{rajbhandari2022deepspeed}.

In summary, the integration of MoE in Discriminative AI significantly broadens the scope and enhances the effectiveness of traditional AI systems. MoE models facilitate a more nuanced and powerful approach to solving complex problems by adopting diverse architectures and ensuring computational efficiency. This seamless merging of MoE with Discriminative AI paves the way for its integration with GAI, particularly in the dynamic and expansive realm of the Metaverse. As we transition to discussing MoE in the context of GAI for the Metaverse, it becomes clear how this synergy can lead to groundbreaking advancements in creating more immersive, interactive, and adaptive virtual environments.

\subsection{Mixture of GAI Expert for the Metaverse}

In the rapidly evolving domain of the Metaverse, the integration of MOE with GAI marks a significant advancement. As shown in Figure \ref{fig:category&applications}, an example advantage of integrating MoE with GAI lies in the enhanced efficiency and scalability. The MoE achieves this by activating only relevant parts of the GAI model necessary for a particular task, thereby managing the Metaverse's vast and complex data more effectively. This aspect is particularly crucial in virtual world generation, where GAI-assisted creation of virtual environments demands handling extensive data and diverse scenarios. For example, the authors in~\cite{fan2022m3vit} illustrated how integrating MoE into a multi-task learning framework can significantly outperform traditional encoder-focused models. This integration not only yields better accuracy in semantic segmentation tasks but also reduces computational resource consumption, highlighting the potent efficiency of MoE models. Additionally, such design enabled high scalability and reduced computational demands which aligns perfectly with the requirements of edge computing, benefiting real-time services by optimizing resource utilization and enhancing operational efficiency. 
\begin{figure*}[t!]
\centerline{\includegraphics[width=\textwidth]{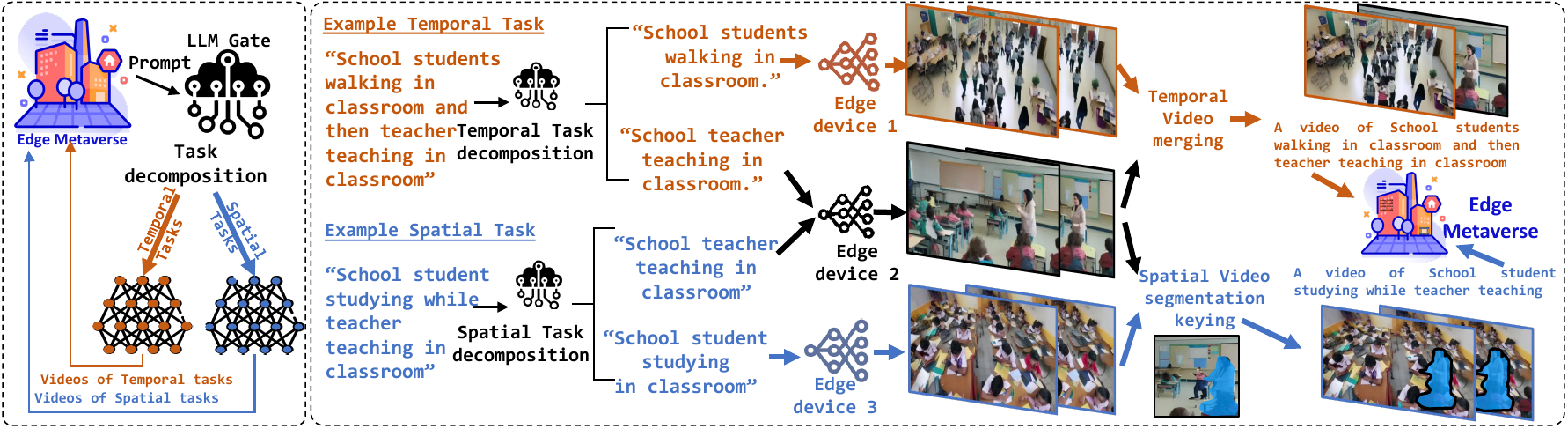}}
    \caption{The mobile edge mixture of video generation framework provides a comprehensive approach to creating video content for the Metaverse. It leverages LLM for task decomposition and expert edge devices for video generation. The framework efficiently handles both temporal and spatial video generation tasks. This strategic division and processing ensure that video content is dynamically generated and merged, emphasizing the collaborative, resource-efficient potential of MoE in creating immersive virtual experiences.}
    \label{fig:framework}
\end{figure*}
Another notable development in this area is the Multi-Modal Variational Autoencoder (MMVAE), which employs an MoE variational posterior across individual modalities~\cite{shi2019variational}. This approach allows for the learning of a generative model capable of handling observations across multiple modalities during training, as well as managing missing modalities at testing. Such an approach is particularly relevant in the Metaverse context, where virtual environments are dynamic and multimodal interactions including visual, audio, and gesture are common. The MMVAE model is uniquely positioned to handle complex many-to-many multi-modal mappings, providing a greater range of flexibility compared to traditional one-to-one image transformations~\cite{shi2019variational}. This model is also pioneering in exploring image-language transformations under the multi-modal VAE setting, a capability crucial for enhancing user experience in the Metaverse.

Moreover, MoE significantly enhances image fusion tasks. The authors in~\cite{cao2023multi} proposed a framework that involves a multi-modal gated mixture of local-to-global experts, where specific tasks are assigned to each expert, enabling sample-adaptive specialized learning. This approach leads to superior performance in dynamically fusing images, an essential capability in the Metaverse where environments are subject to constant change and require high levels of detail and realism. The MoE-Fusion model, by efficiently managing the fusion of multiple modalities, such as infrared and visible light images, significantly enhances the flexibility and realism of virtual environments in the Metaverse. This advancement demonstrates the potential of MoE in augmenting the capabilities of GAI models, thereby contributing to the creation of more immersive and versatile Metaverse services.

In summary, the integration of MoE with GAI for the Metaverse opens up new avenues for creating more efficient, specialized, and dynamic virtual environments. This collaboration is pivotal in driving the future of virtual experiences, making the Metaverse a more immersive, interactive, and responsive platform for users.

\section{Mixture of Video Generation Experts in Metaverse}

In the Metaverse, video generation emerges as a cornerstone for crafting dynamic and customized virtual experiences. This technology enables the creation of virtual environments that dynamically respond to user actions and preferences in real-time to amplify the immersive experience. To address the complexities and resource constraints of video generation in the mobile edge Metaverse, we propose a framework that employs a strategic approach to video content creation by decomposing scenes into subvideos as shown in Figure \ref{fig:framework}, which can be processed separately by different edge servers or devices. This method allows for collaborative and resource-efficient video generation, where tasks can be divided either temporally or spatially.  The proposed framework consists of several key components:

\begin{itemize}
\item \textbf{Task Decomposition Gate} The Metaverse prompt is handled by the fine-tuned LLM gpt-3.5-turbo-0125\footnote{https://platform.openai.com/docs/models/gpt-3-5-turbo} for task decomposition. The original prompt is classified into different tasks such as temporal and spatial video generation tasks and then decomposed into sub-prompts for the edge expert models. In scenarios where the sub-prompt are events change over time, each sub-prompt represents a specific time slice within the overall scene.

\item\textbf{Edge Video Generation} A zero-shot-text-to-video model is deployed as experts for video generation~\cite{Khachatryan_2023_ICCV}. By incorporating motion information into the latent codes of generated frames and employing a novel cross-frame attention mechanism, this model ensures the temporal consistency of the background and preservation of the foreground object's identity throughout the video sequence. Notably, it performs remarkably well even without the need for additional training on video data, thus significantly reducing computational overhead and making high-quality video generation more achievable by edge devices~\cite{Khachatryan_2023_ICCV}. 
\begin{figure*}[t!]
\centerline{\includegraphics[width=0.9\textwidth]{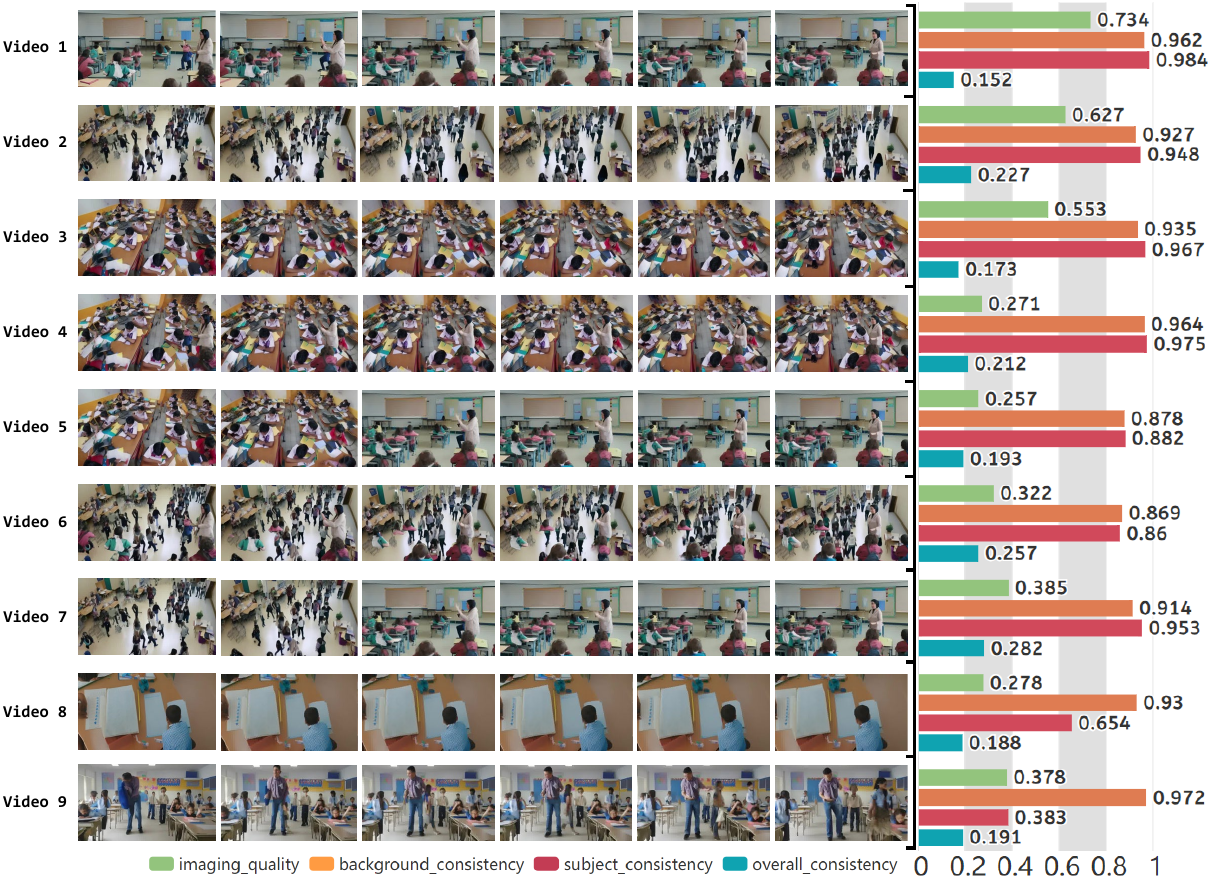}}
   \caption{Videos and respective results. \textbf{Video 1}: ``School teacher teaching in a classroom." \textbf{Video 2}: ``School students walking in classroom." \textbf{Video 3}: ``school student studying in classroom" \textbf{Video 4} is a spatial keying fusion of Video 1 and Video 3, demonstrating the ability to blend different spatial elements into a cohesive scene. \textbf{Video 5} is a temporal merging of Video 1 and Video 3, illustrating how sequences can be combined over time to create a narrative progression. \textbf{Video 6} is a spatial keying fusion of Video 1 and Video 2, highlighting the integration of instructional and dynamic student movement within the same spatial environment. \textbf{Video 7} is another spatial keying fusion of Video 1 and Video 2, further exploring the seamless combination of teaching activities and student interactions. \textbf{Video 8} is generated without our MOE framework using original prompt ``A video of School student studying while teacher teaching".\textbf{Video 9} is generated without our MOE framework using original prompt ``A video of School students walking in classroom and then teacher teaching in classroom".}

    \label{fig:videos}
\end{figure*}
\item\textbf{Video Merging} The video merging component is essential for synthesizing a unified video from the individually generated subvideos generated by experts, utilizing either temporal or spatial strategy based on the classification result by the gating network. Subvideos that are merged temporally are arranged in a sequence that reflects the correct order of events, ensuring logical continuity.  On the other hand, spatial merging employs video segmentation and keying techniques to blend subvideos that occur concurrently but in different spatial areas of a scene, maintaining spatial relationship and coherence. This dual approach guarantees a continuous and dynamic portrayal of the scene. It optimizes resource use by sharing similar subvideos for different tasks, thus avoiding redundant generation efforts and enhancing the efficiency of content creation in the Metaverse.
\end{itemize}

By leveraging MoE, this framework offers a foundational method for mobile edge video generation services to collaborate, sharing resources and optimizing efficiency. The framework not only reduces the load on individual devices that are running diverse expert models, but also speeds up the content creation process, allowing for real-time generation of personalized and immersive virtual experiences in the Metaverse.

\section{Case Studies: MoE in Action within the Metaverse}

In exploring the practical applications of the MoE within the Metaverse, especially for mobile edge video generation services, our case studies illustrate the framework's effectiveness and the quantitative benefits.

\subsection{Evaluation Metrics}
To rigorously evaluate the performance and capabilities of these video generation models, the VBench benchmark suite has been introduced~\cite{huang2023vbench}. VBench is a comprehensive evaluation framework designed to dissect ``video generation quality" into specific, hierarchical, and disentangled dimensions. Each dimension comes with tailored prompts and evaluation methods. This benchmark suite features 16 distinct metrics that assess videos across various aspects, providing a fine-grained analysis of the models' strengths and weaknesses. These dimensions include factors such as subject consistency that measures whether the appearance of the main subject (e.g., a person or an animal) remains consistent throughout the video, imaging quality that measures the distortion (e.g., over-exposure, noise, and blur) presented in each video frame, background consistency that measures whether the background scene remains consistent throughout the video, overall consistency that measures whether the video is consistent with the text prompt in terms of both semantics and style. Thus, VBench’s methodological approach allows for the objective evaluation of video generation models, aligning closely with human perceptions.

\subsection{Evaluation Metrics}
As shown in Figure \ref{fig:videos}, our experiments focused on comparing the effectiveness of MoE structures in different video generation scenarios, particularly examining how subvideos could be combined and shared among edge Metaverse devices. We tested two main types of task decomposition: temporal and spatial, and assessed how accurately these decomposed tasks could be merged to form a coherent and high-quality video final output. In the following results, we employ four evaluation metrics from VBench :
\begin{itemize}
    \item \textbf{Imaging Quality:} Measures the distortion present in generated frames, such as over-exposure, noise, and blur, using the MUSIQ image quality predictor trained on the SPAQ dataset~\cite{huang2023vbench}.
    \item \textbf{Background Consistency:} Assesses the temporal consistency of background scenes by calculating CLIP feature similarity across frames.
    \item \textbf{Subject Consistency:} Evaluates whether the appearance of a subject (e.g., a person, car, or cat) remains consistent throughout the video, using DINO feature similarity across frames.
    \item \textbf{Overall Consistency:} Uses overall video-text consistency computed by ViCLIP on general text prompts to reflect both semantics and style consistency~\cite{huang2023vbench}.
\end{itemize}

These four metrics are computed through distinct models and normalized to a scale of zero to one for average video quality calculation. Moreover, the metric of overall consistency is not derived from subject consistency and background consistency. Overall consistency offers an independent measure that reflects both the semantic and stylistic coherence of the generated videos, underscoring the composite effectiveness of the MoE structures in creating high-fidelity video content. These evaluation metrics are pivotal for our comprehensive assessment of video generation quality in edge Metaverse devices.
\subsection{Results and Discussions}

\begin{figure*}[t!]
\centerline{\includegraphics[width=0.95\textwidth]{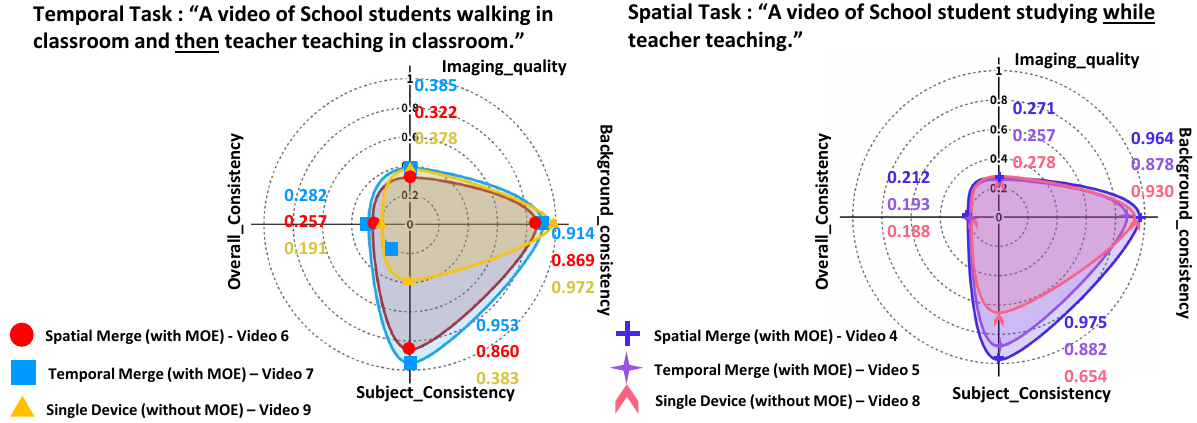}}
   \caption{Results comparison across videos with different merging strategies (temporal and spatial), as well as video generated using one device without the proposed MOE framework.}
    \label{fig:results}
\end{figure*}

Our experiments on the integration of MoE in the Metaverse, mainly focusing on mobile edge video generation, have achieved compelling results that underscore the framework's significant impact on video content creation.  As shown in Figure ~\ref{fig:results}, the average video quality (the average of 4 metrics including imaging quality, background consistency, subject consistency and overall consistency) decreases by 5.65\% due to inappropriate task categorization, such as merging temporal tasks in a manner more suited for spatial tasks, and vice versa. This result underscores the importance of accurate task classification by the gating network and further explains the flexibility of MoE architectures in facilitating diverse video combination methodologies. This inherent adaptability of MoE frameworks ensures the maintenance of a seamless narrative trajectory and visual consistency across video content, effectively navigating the complexities inherent in the disparate elements of the tasks involved.

Moreover, the application of MoE to video combination tasks has markedly improved the visual quality and consistency of the generated videos. By effectively decomposing and then accurately recombining videos, the MoE framework ensures that the resultant content aligns with the intended logic structure and exhibits a high level of coherence. The proposed framework offers a notable improvement over videos generated by a single device using the initial prompt. Specifically, the MoE framework exhibits excellent performance in maintaining subject consistency. When facing complex prompts, single-device video generation methods may overlook secondary or subsequent subjects intended. As shown in Figure~\ref{fig:videos}, it is observed that Video 8 (generated by a single device with the entire prompt) does not contain the ``Teacher" subject. As illustrated in Figure~\ref{fig:results}, the subject consistency of video 8 is 32.1\% lower than that of Video 4 (generated by the MoE framework).

In other evaluative metrics, videos produced under the MoE framework also demonstrate enhanced performance relative to those generated without it, except for background consistency. The cause could be that merged videos tend to be more complex than single videos, potentially affecting background uniformity. Nonetheless, regarding overall consistency, the MoE framework significantly outperforms single device, underlining its effectiveness in improving the logic of video content. 

Quantitatively, the improvement validated by the VBench benchmark suite employing MoE structures was evident across a range of metrics. Videos recombined in alignment with their specific decomposition type demonstrated great consistency and imaging quality.

\section{Future Directions}

While the advantages of MoE are evident, it is also important to consider the challenges that come with its implementation. In practical applications, a balance must be tackled between various performance aspects and parameter trade-offs. 
\begin{itemize}
    \item \textbf{Training Complexity:} Training MoE models are notably more complex and resource-intensive due to their additional gating mechanisms. This complexity can be mitigated by employing parallel computing resources and simplifying the gating mechanism.
    \item \textbf{Expert model Design:} The selection and design of expert models within an MoE framework are crucial for its overall performance. The challenge lies in balancing the diversity and efficiency of experts. For Mobile edge Metaverse, selection of local devices and grouping under different gating networks could be a possible future direction.
    \item \textbf{Communication Bandwidth Bottlenecks:} MoE frameworks used in distributed computing settings often encounter issues related to communication bandwidth, especially when it comes to data traffic exchanged between experts. To address this bottleneck, it is essential to design an architecture that minimizes unnecessary data transfers.
\end{itemize}

\section{Conclusion}
Integrating MoE with GAI within the mobile edge Metaverse can significantly open new avenues for immersive and dynamic virtual experiences. Our exploration reveals that this integration remarkably addresses the existing challenges in content creation and indicates a new era of personalized and adaptive virtual environments. We have demonstrated improved video content quality and consistency through our proposed framework and case studies. However, challenges such as training complexity, expert model design, and communication bandwidth remain, which can be further refined to fully harness the potential of MoE and GAI in the Metaverse. 

% \newpage
\bibliographystyle{IEEEtran}
\bibliography{main}
\end{document}